\documentclass[utf8]{FrontiersinHarvard} 

\usepackage[T1]{fontenc}
\usepackage[utf8]{inputenc}

\usepackage{tabularx}
\usepackage{booktabs}
\usepackage{caption}
\usepackage{tabularx}
\usepackage{graphicx}
\usepackage{adjustbox}

\usepackage{xspace}

\usepackage{url,hyperref,lineno,microtype}
\usepackage[onehalfspacing]{setspace}

\newcommand{\asaga}[0]{aSAGA-UA\xspace}

\newcommand{\psg}{PSG\xspace}
\newcommand{\psgs}{PSGs\xspace}

\def\keyFont{\fontsize{8}{11}\helveticabold }
\def\firstAuthorLast{Holm {et~al.}} 
\def\Authors{
Benedikt Holm\,$^{1,2,*}$, 
Gabriel Jouan\,$^{1,2,*}$,
Emil Hardarson\,$^{1,2}$,
Sigríður Sigurðardottir\,$^{2}$,
Kenan Hoelke\,$^{2,4}$,
Conor Murphy\,$^{2,3}$,
Erna Sif Arnardóttir\,$^{1,2}$,
María Óskarsdóttir\,$^{1,2}$,
and Anna Sigríður Islind\,$^{1,2}$}
\def\Address{$^{1}$Reykjavik University, Department of Computer Science,  Reykjavik, Iceland \\
$^{2}$Reykjavik University Sleep Institute, Reykjavik, Iceland \\
$^{3}$Physical Activity, Physical Education, Sport and Health Research Centre (PAPESH), Sports Science Department, School of Social Sciences, Reykjavik University, Reykjavik, Iceland\\
$^{4}$Board of Registered Polysomnographic Technologists, Arlington, VA, USA }
\def\corrAuthor{Corresponding Author}

\def\corrEmail{benedikthth@ru.is}

\begin{document}
\onecolumn
\firstpage{1}

\title[Sleep Revolution Data Platform]
{An Optimized Framework for Processing Large-scale Polysomnographic Data Incorporating Expert Human Oversight} 

\author[\firstAuthorLast ]{\Authors} 
\address{} 
\correspondance{} 

\extraAuth{}

\maketitle

\begin{abstract}

\section{}
Polysomnographic recordings are essential for diagnosing many sleep disorders, yet their detailed analysis presents considerable challenges. With the rise of machine learning methodologies, researchers have created various algorithms to automatically score and extract clinically relevant features from polysomnography, but less research has been devoted to how exactly the algorithms should be incorporated into the workflow of sleep technologists.  
This paper presents a sophisticated data collection platform developed under the Sleep Revolution project, to harness polysomnographic data from multiple European centers. 
A tripartite platform is presented: a user-friendly web platform for uploading three-night polysomnographic recordings, a dedicated splitter that segments these into individual one-night recordings, and an advanced processor that enhances the one-night polysomnography with contemporary automatic scoring algorithms.
The platform is evaluated using real-life data and human scorers, whereby scoring time, accuracy and trust are quantified. Additionally, the scorers were interviewed about their trust in the platform, along with the impact of its integration into their workflow.

\tiny
 \keyFont{ \section{Keywords:} Sleep Research, platform, Human-In-The-Loop, Machine Learning, Scoring time, Agreement, Explainable AI, trust} 
\end{abstract}

\section*{Introduction}
    The emergence of explainable artificial intelligence (XAI) presents vast potential for revolutionizing various application areas, such as in healthcare \citep{de2023explainable}.
    However, despite the great potential, there are significant issues that need to be tackled before XAI can be fully utilized \citep{jermutus2022influences}. 
    One such issue originates from application areas within healthcare, where automation of manual tasks and data-driven decision-support has to take the central stage before XAI can become a viable option \citep{loh2022application}.

    A subfield of healthcare is the collection and analysis of sleep recordings, referred to as polysomnography~(PSG) \citep{arnardottir2021future}.
    A PSG is an overnight recording of various biomedical signals, such as an electroencephalogram~(EEG), electromyogram~(EMG), electrooculogram~(EOG), and various respiratory signals. Upon collection, the PSG must be manually annotated by a sleep technologist which is a cumbersome and time-consuming task \citep{arnardottir2021future}.
    PSG scoring is a vital step in the process of identifying and diagnosing the presence of many sleep disorders, some of which are extremely prevalent \citep{Benjafield2019-gh}.
    A sleep technologist will manually review the recording according to a set of rules devised by the American Academy of Sleep Medicine (AASM), labeling events such as respiratory events, and sleep stages in a process referred to as scoring. The sleep stage scoring is done by assigning a sleep stage to each 30-second segment (also called epochs) in the recording. 
    A product of the PSG scoring is the creation of a hypnogram, a graphical representation tracing the progression of sleep stages throughout the night. This visual tool, often complemented by a hypnodensity graph, provides a detailed overview of the patient's sleep architecture, capturing transitions between sleep stages \citep{jang2022recurrent,Pevernagie2024-nj}. 
    Self-applied-PSG (henceforth referred to as simply \psg), a newly designed simplified version of traditional PSG, utilizing frontal EEG instead of the conventional International 10-20 System, refers to a type of sleep study that the participant can set up themselves and sleep with at home for up to three nights in the current work \citep{arnardottir2022sleep}.

    One of the main drawbacks of the current scoring process is, as stated earlier, that it can be excessively time-consuming, which can cause considerable delays in providing sleep reports to healthcare providers and consequently delay diagnosis \citep{biedebach2023}, as well as increase the cost of healthcare considerably \citep{wickwire2021there}. Adding to this challenge, significant inter-scorer variability exists \citep{Nikkonen2024-km}; disagreements can reach  19.3\% for sleep stages \citep{Nikkonen2024-km} and 11.6\% for respiratory events \citep{doi:10.5664/jcsm.26818}. Delays and disagreements such as these can have negative effects on patient outcomes, as untreated sleep disorders can have a significantly negative impact on patient health \citep{Dikeos2011-tr}.
    
    The advent of machine learning and other automatic scoring algorithms offers a potential solution by automating the process of manual scoring, which the AASM sees great potential in \citep{goldstein2020artificial}. However, the development and application of machine learning are often prohibitively technical, requiring diverse knowledge of computer science to achieve \citep{Giray2021-us, brennan2022barriers}. 
    There is also a dire need for socio-technical alignment, i.e. the multi-disciplinary collaboration between the computer scientists integrating the algorithms, and the professionals working in the context in which the algorithms are being integrated \citep{brennan2022barriers}. The integration of AI, machine learning, or advanced data-driven decision-making of any kind into the workflow may move the industry professionals from a generative role (creating the outputs themselves) to the role of auditors, where they correct the output of the algorithms, and consult with computer scientists to tweak and alter the models to handle edge cases or incorrect generations by the algorithm \citep{Gronsund2020-te}.
    Moreover, in the rare case when socio-technical alignment is reached, trust issues often surface, where the professionals working within the context that the algorithms are integrated into, may not trust the outcomes \citep{mistrust}, which has posed a great limitation in healthcare \citep{jermutus2022influences, lee2021application}. This mistrust has received limited focus in terms of research contributions and needs to be studied further.

    Machine learning models are often deemed a 'black box,' owing to their lack of transparency and the extensive technical knowledge needed to understand them. Moreover, their incapacity to adapt to dynamically evolving requisites often leads to their obsolescence. This has resulted in the increasing prevalence of human-in-the-loop AI systems \citep{Mosqueira-Rey2023}. Human-in-the-loop AI systems allow one or more human experts to take an active part in the training process by continuously evaluating the model and providing new inputs that are then selectively used to re-train the model in a process called active learning \citep{settles2009active}.

    To advance and modernize sleep research as well as to enable the collection of a large-scale European sleep recording dataset, the Sleep Revolution project, a joint venture involving 24 European partners, was initiated \citep{arnardottir2022sleep}. 
    Each partner contributes approximately 60 sets of three-night \psgs. Sleep technologists then evaluate these on a shared workstation which is a part of the Sleep Revolution high-performance cluster. After this, healthcare professionals analyze sleep parameters, which helps them to diagnose the patient.
    A significant objective of the Sleep Revolution is to reduce scoring time \citep{arnardottir2022sleep}. One strategy to achieve that goal is to direct the focus of the sleep technologists to the areas of sleep that automatic algorithms have less 'certainty' of. By displaying these areas of high uncertainty, referred to as \emph{gray areas} from now on, we can specifically target the sleep technologists towards these areas, instead of unilaterally trusting or mistrusting the automatic scoring algorithms \citep{jouan2023algorithmic}. 
    
    To enable these algorithms to benefit sleep technologists in their daily work, a system is required that bridges the gap between the data collection and the manual scoring itself. To collect the data required for this work, a digital platform was designed to handle automatically collecting, segmenting and processing the PSG. 
    The concept of digital platforms takes into account that a digital platform is both a piece of software, while it is also an intermediary that connects needs with resources. Therefore the concept of digital platforms encompasses a larger array than the software itself as it, in a socio-technical manner, also takes the context into account. 
    In this case, the digital platform is accessed via the users' web browser and is hereinafter referred to as the platform.

        Computer-assisted automatic scoring with manual review has demonstrated the ability to reduce PSG scoring time significantly, with some studies showing improvements by factors of 1.26 to 2.41\citep{Alvarez_Estevez2022_by}. Moreover, automatic sleep scoring algorithms can halve the scoring time \citep{Liang2019-st, chooBenchmarking}.

        Some research on the integration of automated scoring has been conducted in the last few years as listed in Table \ref{tab:litcomparison}. \citet{10.1093/sleep/zsad275} discuss the challenges and advancements in automatic sleep scoring in the context of rodent and human sleep research. They note limitations in handling atypical data and lack of flexibility but also note that automatic algorithms can make the process more efficient. 
        A recent study evaluated a deep-learning-based automatic scoring software for its accuracy and efficiency compared to manual scoring. The results indicated a high correlation between the automatic scoring system and manual scoring, particularly in sleep staging and the apnea-hypopnea index. The automatic scoring system also demonstrated a significant reduction in manual scoring time, leading to improved workflow efficiency in sleep laboratories \citep{10.3389/fneur.2023.1123935}. 
        \citet{info:doi/10.2196/17971} interviewed 9 healthcare professionals and 5 patients about their attitudes towards using data from electronic health records in an algorithm to screen for alcohol abuse in hospitals. Professionals were mixed in their views, appreciating the tool's time-saving potential but concerned about losing instinctual decision-making. While this work is only tangentially related to our work, the authors point out the requirement to include healthcare professionals in the process of integrating automatic algorithms. 
        \citet{Gerla2018-pz} presented a computer-assisted approach for sleep staging using EEG recordings and AASM 2012 scoring rules, focusing on real clinical data with artifacts and missing electrodes,  evaluating the influence of AI in clinical settings by comparing traditional manual sleep stage classification with AI-based methods, including expert-in-the-loop strategies, for the analysis of EEG recordings in sleep studies.
        In a later study, \citet{Gerla2019-si} developed a semi-supervised method for evaluating PSG, blending expert-scored segments with automated classification. This approach, tested on both healthy individuals and chronic insomnia patients, showed enhanced efficiency and accuracy in sleep data analysis compared to conventional manual scoring methods, demonstrating the impactful role of AI in streamlining sleep study workflows.

        \begin{table}[h]
            \centering
            \begin{tabularx}{\linewidth}{l >{\centering\arraybackslash}X >{\centering\arraybackslash}X >{\centering\arraybackslash}X}
                \toprule
                \textbf{Work} & \textbf{\small{Addresses integration into existing work environments}} & \textbf{\small{Addresses impact on workflow}} & \textbf{\small{Addresses opinions of medical professionals on AI in workflow}} \\
                \midrule
            
                Our work & X & X & X \\
                \citet{10.1093/sleep/zsad275}       &   &X  &   \\
                \citet{10.3389/fneur.2023.1123935}  &   &X  &   \\
                \citet{info:doi/10.2196/17971}      &   &   &X  \\
                \citet{Gerla2018-pz}                &   &X  &   \\
                \citet{Gerla2019-si}                &   &X  &   \\
                \bottomrule
            \end{tabularx}
            \caption{Comparison of contributions of this work and similar work.}
            \label{tab:litcomparison}
        \end{table}

    \subsection*{Contributions}
As is evident from Table \ref{tab:litcomparison}, existing research on automatic sleep scoring addresses either the impact on workflow or the opinions of medical professionals on AI in the workflow.
To the best of our knowledge, no research exists that addresses the integration of automatic sleep scoring into existing work environments which is an important aspect to consider to achieve socio-technical alignment. 

To fill this research gap, we designed both a platform and a process for evaluating the effectiveness of introducing gray areas into the work of sleep technologists and their trust in the process.
By integrating the platform featuring machine learning algorithms into the work of sleep technologists through our empirical case within the Sleep Revolution, we extrapolate three main contributions. 
Firstly, we outline the architecture for a platform that has been designed and developed to enable the integration of automatic scoring. 
Secondly, we introduce the concept of "gray areas" as a method of selectively focusing the attention of sleep technologists on fewer areas in the PSG.
Thirdly, we illustrate the decreased scoring time and increased agreement gained by integrating the automatic scoring algorithms into the workflow of sleep technologists.
Throughout this research, and particularly when analyzing the results, we realized that the phenomena we encountered consistently and that was common to all of our results, was missing a clear clinical terminology that we attempt to address in this work.

\section*{Materials and Methods}
    PSG sharing and scoring between research centers require sophisticated architectures that rest heavily on the principles of storing and processing medical data cohesively. The proposed platform has the main purpose of connecting needs with resources, which in this case outlines the sharing and scoring of PSG between research centers. 
     
    The methodology is three-fold; 1) the design and development of the platform, 2) the validation of the platform, and 3) interviews with sleep technologists. The design section covers the architecture, components, and technologies chosen to implement the platform, the validation section covers how the platform was assessed in terms of processing duration, sleep technologist speed, and agreement improvements, and the interview section describes how sleep technologists were interviewed for their sentiment toward integrating AI tools into the workflow.
    
    \subsection*{Platform design}
    
        The platform needed to be conceived in agreement with the main constraints as having a simple user interface for sleep technologists to be able to authenticate and upload their \psg; providing administrative oversight on uploads from different centers; being fault-tolerant; and being scalable \citep{noauthor_undated-tv}. The platform is split into three distinct components:
        \begin{enumerate}
            \item Web-based front-end for user uploads, administration, and dispatching of jobs to the other components (henceforth referred to as the \textit{front end}).
            \item Three-night \psg splitter (henceforth referred to as the \textit{splitter}).
            \item Processing pipeline that augments \psgs with automatic scorings (henceforth referred to as the \textit{processor}).
        \end{enumerate}
        
\begin{figure}[h]
            \centering
            \includegraphics[width=\linewidth]{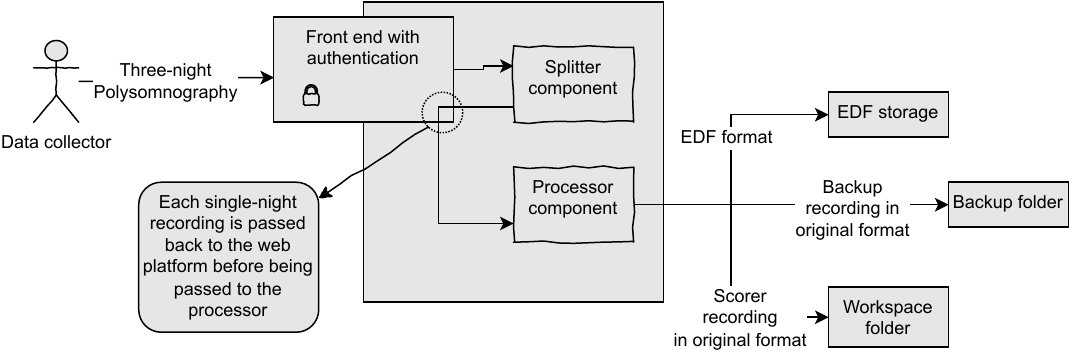}
            \caption{Overview of the platform showing how the front end, processor, and splitter are combined.}
            \label{fig:highlevelview}
        \end{figure}

        An overview of the platform architecture is shown in Figure \ref{fig:highlevelview}. An important feature of the platform is to allow users (e.g. sleep technologists and healthcare professionals) to upload multiple \psgs to be shared and scored at the same time without breaking the platform. To achieve this, the FastAPI Python web framework was used, which despite its simplicity handles multi-user web applications supporting asynchronous code \citep{fastapi}.
        The platform is protected with a user login access in which each user is a validated member of Sleep Revolution consortium\citep{SleepRevoGrant}.
        
        Additionally, the front end handles receiving signals from both the splitter and the processor via HTTP requests, as well as issuing jobs to the splitter when a new \psg is received, and to the processor when a \psg has successfully been split. The splitting is a necessary step when several nights' PSG are combined into one file.  
        The job queue was achieved using a RabbitMQ queuing server, which is a program that allows disparate asynchronous programs to communicate by listening and issuing messages to a queue \citep{rabbitmq}.
        By utilizing a message-queue protocol, the font end can offload more time and memory-consuming projects such as generating automatic scorings to other processes, thus reducing the probability of users experiencing downtime, or data loss.
        
        The processor is the final component of the architecture. Its purpose is to prepare the individual night \psg by augmenting the \psg with the AI scoring, along with the gray area scoring. The output of the processor is twofold. Firstly, the processor prepares a 'scoring' version of the \psg that is augmented with predicted sleep stages from an automatic scoring algorithm integrating gray areas and is made available for manual scoring, and a version meant for later computer processing and machine learning.
        Each component was containerized using the virtualization software Docker \citep{merkel2014docker} for enhanced isolation, consistency, and reproducibility during deployments, which is important in sustainable and secure development.

        As introduced previously, the processor prepares the \psg to be manually scored, stored, and ready for further analysis. To reduce the manual work of the sleep technologists, a crucial step in the processor is highlighting areas in the \psg that are hard to score for the algorithm, i.e. gray areas. 
        The gray area augmentation works first by sending each one-night \psg EDF file to the trained deep learning model aSAGA \citep{rusanen2023asaga}. 
        The aSAGA architecture is based on a revisited U-time architecture for scoring and respiratory events prediction \citep{perslev2021u,huttunen2022comparison}.
        The U-time is an encoder-decoder structure consisting of blocks of consecutive convolutional, batch normalization, and pooling layers. 
        However, in the aSAGA algorithm, a single-channel model is used, which was first trained on PSGs' EEG (C4-M1) and then fine-tuned with an EOG (E1-M2) channel using self-applied \psgs with frontal setup. 
        This was done to have generalizability between EEG and EOG channels and to increase the compliance of the model for frontal EEG and EOG setups.
        The aSAGA model is parameterized to return a hypnogram of the same length as the number of epochs from the signal input. 
        The model has an accuracy of 80\% estimated over different scored sleep datasets. This accuracy is on par with manual scoring \cite{Nikkonen2024-km}, however, the gray areas from aSAGA model prediction have been validated by comparing the match with the gray areas from predicted manual scoring uncertainty.

        \begin{subfigure}[h!]
            \setcounter{figure}{2}
                \setcounter{subfigure}{0}
                \begin{minipage}[b]{0.95\linewidth}
                    \includegraphics[scale=0.54]{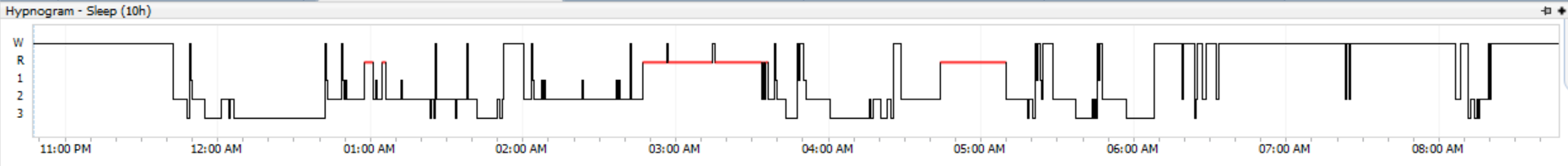}
                    \caption{aSAGA predicted hypnogram.}
                    \label{fig:aSAGA}
                \end{minipage}
            
            \setcounter{figure}{2}
                \setcounter{subfigure}{1}
                \begin{minipage}[b]{0.95\linewidth}
                    \includegraphics[scale=0.54]{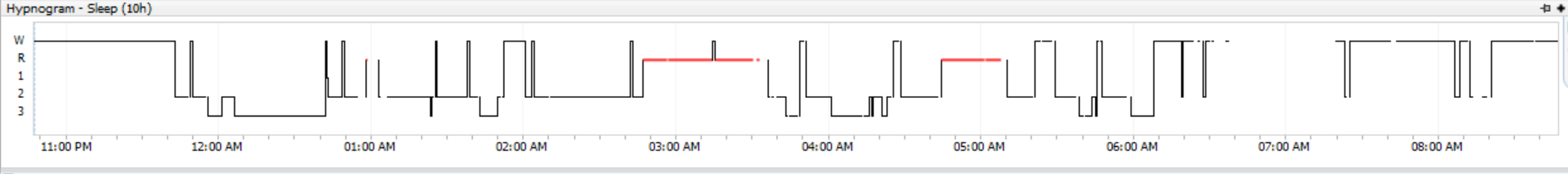}
                    \caption{\asaga predicted hypnogram with gray areas as whitespace.}
                    \label{fig:aSAGAun}
                \end{minipage}
            \setcounter{figure}{2}
            \setcounter{subfigure}{-1}
            \caption{Example of output hypnograms for the n°1 \psg from $50\times 10$\psg, obtained using the processor and rendered in Nox Medical's Noxturnal software for manual review.}
            \label{fig:NOXVIEW}
        \end{subfigure}

        The second part concerns the gray areas. 
        Using the predicted hypnodensity from the aSAGA model as input, a trained clustering algorithm tags each epoch that belongs to the gray areas \citep{jouan2023algorithmic}. 
        The clustering algorithm is a multi-objective method based on multinomial mixture models clustering the different levels of sleep technologist agreement and summarizing the results into two sets of high-agreement and \emph{gray area} clusters. 
        The threshold is selected according to the maximization of the distance between two distributions of the sleep technologist's agreement measure. 
        When the algorithm receives a new hypnodensity, it outputs a hypnogram called \asaga with "whitespace" epochs for the gray areas instead of regularly scored epochs. 
        Figure \ref{fig:aSAGAun} gives an example of a rendered hypnogram with gray areas. 
        
        Using \asaga, it becomes easier for the sleep technologists to view epochs where the AI scoring may not be accurate, and need to be re-evaluated. 
        For instance, in Figure \ref{fig:aSAGA} between 00:50 a.m. and 1:10 there are many transitions between N2, N1, and REM scored by the algorithm.
        In Figure \ref{fig:aSAGAun}, it is clear that the same period is full of gray areas. Regarding the high number of sleep transitions happening in a few minutes, the associated signal might be hard to interpret by the algorithm. A manual review from the sleep technologist is needed in that part of the hypnogram.
        The method has been evaluated on a real case of uncertainty analysis of 50 \psgs manually scored by 10 sleep technologists. We refer to this dataset as $50\times 10$\psg. This dataset comes from a cohort of 50 participants that have previously been scored by ten independent sleep technologists to create a consensus scoring \citep{jouan2023algorithmic,rusanen2023asaga}. 
        After testing the clustering algorithm on predicted hypnodensities from aSAGA, the threshold separating the gray area clusters from other epochs was lowered to $0.73$ according to a sleep technologist's recommendations. 
        The new value avoids the creation of an excessive amount of white spaces in the final hypnogram.

        \subsection*{Sleep Technologist Time and Consensus Validation}
            The platform is validated with the help of three sleep technologists, referred to from this point as Sleep Technologist One, Sleep Technologist Two, and Sleep Technologist Three (ST1, ST2 and ST3 respectively). ST1 and ST2 are experienced sleep technologists, whereas ST3 is considered less experienced.
            Each sleep technologist was asked to score a randomly selected subset of from the $50\times 10$\psg. 
            Each sleep technologist received half of the subset scored with a default proprietary industry-standard automatic scoring and the other half had the automatic scoring with gray areas (\asaga). We also refer to these to options as without and with AI, respectively. The partitioning of the subsets can be seen in Table \ref{tab:GA}.

            \renewcommand{\arraystretch}{1.5}
            \begin{table}[h]
               \begin{center}
                \begin{adjustbox}{center, width=\columnwidth-10pt}
                \begin{tabular}{|c|c|c|c|c|c|c|c|c|c|c|}
                \hline
                \psgs & 1 & 2 & 3 & 4 & 5 & 6 & 7 & 8 & 9 & 10\\
                \hline
                ST1 & X&O&X&O&X&O&X&O&X&O\\
                ST2 & O&X&O&X&O&X&O&X&O&X\\
                ST3 & O&X&X&X&X&X&O&O&O&O\\
                \hline
                \end{tabular}
                \end{adjustbox}
                \end{center}
                \caption{The layout of \psgs to be scored, where X indicates default automatic scoring and O indicates \asaga, that is aSAGA with gray areas. The numbers correspond to specific recordings in the $50\times 10$ \psg.}
                \label{tab:GA}
            \end{table}
            
            The sleep technologists were instructed to score sleep stages and arousals. 
            The \psgs with the default automatic scoring were manually reviewed as sleep technologists would normally do in a clinical setting, reviewing every epoch manually. For the \psgs with  \asaga, only the gray areas were manually reviewed by the sleep technologists. The standard operating proceeding follows these specific steps:
            \begin{enumerate}
                \item Start by running automatic analysis.
                \item Adjust the time frame from lights out to lights on (start and stop times for the correct analysis period).
                \item Score sleep stages and arousals according to AASM version v. 3.0 \citep{Troester_aasm_2023}.
                \item For the \asaga scoring, after reviewing all visible gray areas, look for possible missed epochs by searching for sleep stage scorings that contain the word "uncertain", and correct them. 
            \end{enumerate}

             Each sleep technologist was asked to accurately measure the duration of the scoring process for each \psg in their subset. Subseqently, their scoring was collected andcompared it to the existing consensus scoring from the $50\times 10$ \psg.  The scoring accuracy of the sleep technologists using the system as support was assessed via Fleiss's multi-rater \citep{Fleiss1971-yg} $\kappa$ coefficient. This coefficient $\kappa \in [0,1]$ measures the agreement of the current sleep technologist sequence to the scoring sequences given by the ten sleep technologists in the consensus scoring. In the case of samples with high agreement between sleep technologists, Fleiss's $\kappa$ coefficient converges to 1 and 0 otherwise.

        \subsection*{Interviews with Sleep Technologists}
                
            The perceived trust and reliability of the automated scoring system were evaluated through semi-structured interviews with the three sleep technologists, following an interview guide. These 30-minute interviews aimed to explore the sleep technologists' confidence in the system's output and their comfort in integrating the system into their workflow. The sleep technologists provided feedback on the system's overall performance, as well as reflected on their trust in the system's automatic scoring algorithm and gray area identification.
            The interviews were transcribed verbatim and relevant segments of the interviews were and the qualitative data was analyzed with thematic analysis.

\section*{Results}
    
    This section is divided into three main subsections. 
    Firstly, we present the performance of the platform itself.
    Secondly, we present the performance gain in terms of both scoring time and agreement of the sleep technologists. 
    Thirdly, we present the results from interviews with sleep technologists. 

\subsection*{Platform Performance}
    
    \renewcommand{\arraystretch}{1.5}
    \begin{table}[h]
       \begin{center}
        \begin{adjustbox}{center, width=\columnwidth-50pt}
        \begin{tabular}{|c|c|c|c|}
        \hline
        \multicolumn{2}{|c|}{Splitter}&\multicolumn{2}{c|}{Processor}\\
        \hline
        File Size (Mo) & Processing time (min) & File Size (Mo) & Processing time (min)\\
        \hline
        $1920$  & $1.8 \pm 0.3$ & $640$ & $3.4 \pm 0.3$\\
        $2160$  & $2.1 \pm 0.0$ & $720$ & $4.6 \pm 0.3$\\
        $2400$  & $2.0 \pm 0.0$ & $800$ & $4.8 \pm 1.2$\\
        \hline
        \end{tabular}
        \end{adjustbox}
        \end{center}
        \caption{Samples of processing time in minutes (min) taken by each queue according to file size in mego octets (Mo).}
        \label{tab:parametersSimu}
    \end{table}
    
    Table \ref{tab:parametersSimu} lists the time taken by the two main components of the pipeline, the splitter and the processor.
    Since the platform must split the upload into individual nights before further processing, the initial three-night \psg takes approximately 458.5 seconds (7.6 min) to become available for scoring, including both splitting and processing time. 
    However, for the subsequent \psgs, the sleep technologists mainly perceived the processing time, which averages 336.2 seconds (5.6 min) per \psg.
    The processing time for later \psgs is negligible, as the sleep technologist can begin scoring the first file while the others are being processed. Consequently, the processing time is optimally utilized, preventing any significant delays in the scoring workflow.
    When employed in the early stages of the data collection, the queue sizes of the splitter and processor did not grow to excessive lengths, with the processor queue generally not exceeding the size of three pending processing jobs.

    \subsection*{Sleep Technologist Time and Consensus Validation}

        This section provides insights into sleep technologists' performance when using the AI for scoring. Each sleep technologist received an identical set of \psgs to score, both with and without \asaga. Here, the analysis encompasses the sleep technologist's time efficiency and agreement metrics.
        
        \begin{subfigure}[h!]
                \setcounter{figure}{3}
                \setcounter{subfigure}{0}
                    \centering
                    \begin{minipage}[b]{0.45\linewidth}
                    \includegraphics[scale=0.5]{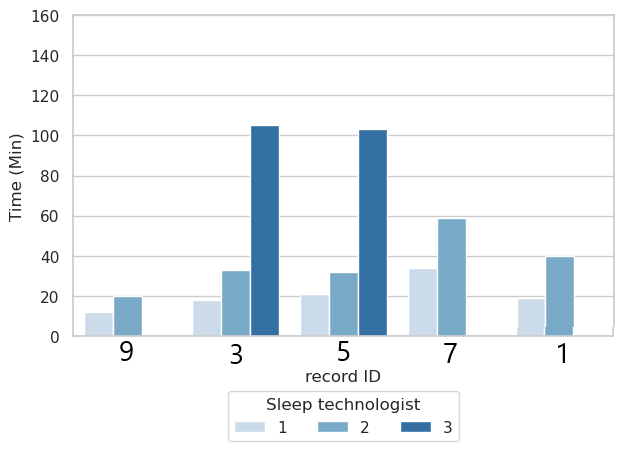}
                        \caption{Scoring duration with \textbf{ST1} using pipeline \asaga assistance.}
                        \label{fig:TS1}
                    \end{minipage}
                    \hfill
                    \setcounter{figure}{3}
                    \setcounter{subfigure}{1}
                    \begin{minipage}[b]{0.45\linewidth}
                    \includegraphics[scale=0.5]{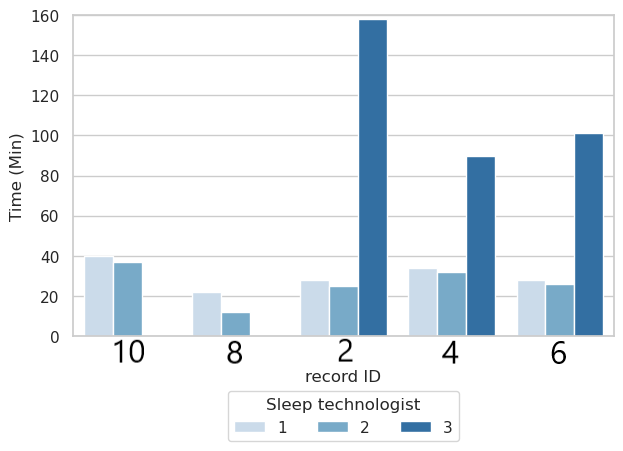}
                        \caption{Scoring duration with \textbf{ST2} using pipeline \asaga assistance.}
                        \label{fig:TS2}
                    \end{minipage}
                     \hfill
                    \setcounter{figure}{3}
                    \setcounter{subfigure}{2}
                    \begin{minipage}[b]{0.45\linewidth}
                    \includegraphics[scale=0.5]{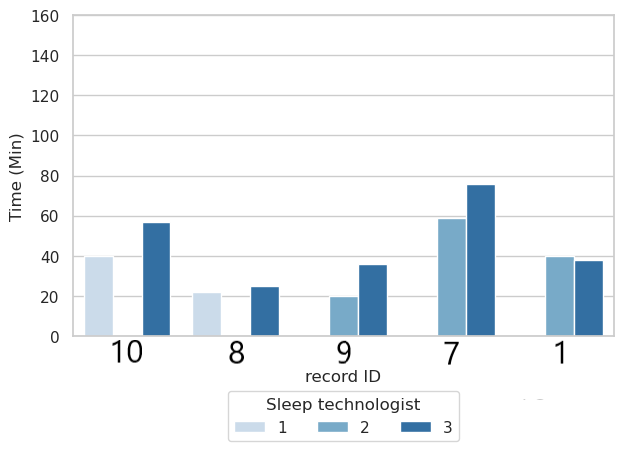}
                        \caption{Scoring duration with \textbf{ST3} using pipeline \asaga assistance.}
                        \label{fig:TS3}
                    \end{minipage}
                \setcounter{figure}{3}
                \setcounter{subfigure}{-1}
                \caption{Scoring duration comparison of \psg with one sleep technologist using \asaga and the other two using the standard procedure.}
                \label{fig:TS}
            \end{subfigure}
            
            In the following section, the study results first delve into the outcomes of the sleep technologist's time to score. Figure \ref{fig:TS} displays the scoring duration for each of the 10 \psgs and all sleep technologists with and without \asaga assistance. 
            As seen in Fig. \ref{fig:TS1}, ST1 using \asaga assistance shows an average~$\pm$~standard deviation scoring duration of 20.8$\pm$8 min compared to 36.8$\pm$16 min for ST2. 
            ST1 reviews faster than ST2. This efficiency translates into an average scoring duration reduction of 16 minutes.
            
            Meanwhile, as seen in Figure \ref{fig:TS2}, when using \asaga, ST2 approximately equaled the time of ST1. 
            ST2 displayed a scoring duration of 26$\pm$9 min, and ST1 displayed 30$\pm$6 min, with ST2 reducing their mean scoring duration by 4 minutes when using \asaga. 
            Finally, Figures \ref{fig:TS1} and \ref{fig:TS2} display ST3 having a time of 111$\pm$26.7 min without AI. 
            ST3, as the least experienced in this study, was noticeably slower in scoring than the other sleep technologists.
            However, Figure \ref{fig:TS3} shows that ST3 depicted a significant decrease in the time to score, of 46$\pm$20.2 when using \asaga, or a reduction of 65 minutes.

            Turning to the agreement analysis, Table \ref{TAB:KAPPA} is divided into two parts; the first half details the sleep technologist's agreement based on the analysis of the complete \psgs, while the second half assesses the agreement specifically for the gray area epochs which the sleep technologist handled with \asaga assistance.
            
            \begin{table}[h]
            \centering
                \begin{tabular}{|c||cc|cc|}
                    \hline
                     & \multicolumn{2}{c}{Complete hypnogram} & \multicolumn{2}{|c|}{Gray areas only}\\
                    \cline{2-5}
                    Sleep Technologist & Without AI & With AI & Without AI & With AI \\
                    \hline
                    \hline
                    ST1 & 0.87$   \pm$0.05 & 0.86$\pm$0.05 & 0.76$\pm$0.12 & 0.73$\pm$0.08 \\
                    \hline
                    ST2 & 0.72$\pm$0.08 & 0.85$\pm$0.04 & 0.48$\pm$0.22 & 0.65$\pm$0.17 \\
                    \hline
                    ST3 & 0.84$\pm$0.04 & 0.80$\pm$0.08 & 0.60$\pm$0.11 & 0.77$\pm$0.12  \\
                    \hline
                \end{tabular}
                \caption{Fleiss's multi-rater $\kappa$ mean$\pm$standard deviation estimated on overall hypnograms and gray areas epochs only by sleep technologists manually scoring and using \asaga assistance}
                \label{TAB:KAPPA}
            \end{table}
            
            When the agreement was calculated based on gray areas, the sleep technologist using \asaga assistance was the only one aware of the nature of these epochs. A steady trend in the overall agreement of sleep technologists using AI for scoring is observed in Table \ref{TAB:KAPPA}. However, the agreement rating of ST3 appears to be negatively affected by the use of \asaga assistance. This reduction is possibly attributed to a more challenging sample of associated \psgs, which generally achieved a lower agreement score, but this is not clear.

            \begin{subfigure}[h!]
                \setcounter{figure}{4}
                \setcounter{subfigure}{0}
                    \centering
                    \begin{minipage}[b]{0.445\linewidth}
                    \includegraphics[scale=0.45]{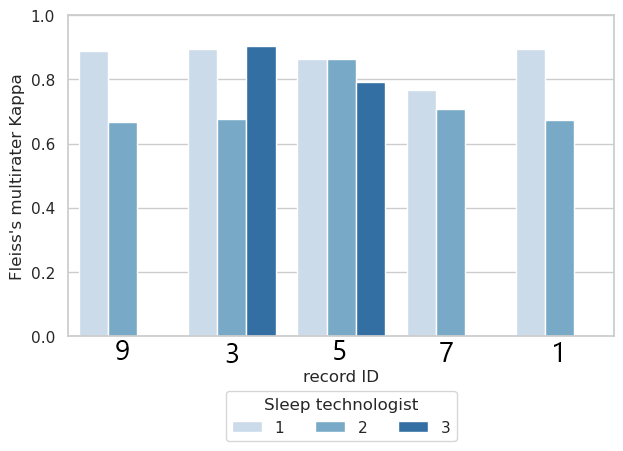}
                        \caption{Fleiss's multiraters $\kappa$ overview with \textbf{ST1} using \asaga assistance.}
                        \label{fig:FKO1}
                    \end{minipage}
                    \hfill
                    \setcounter{figure}{4}
                    \setcounter{subfigure}{1}
                    \begin{minipage}[b]{0.445\linewidth}
                    \includegraphics[scale=0.45]{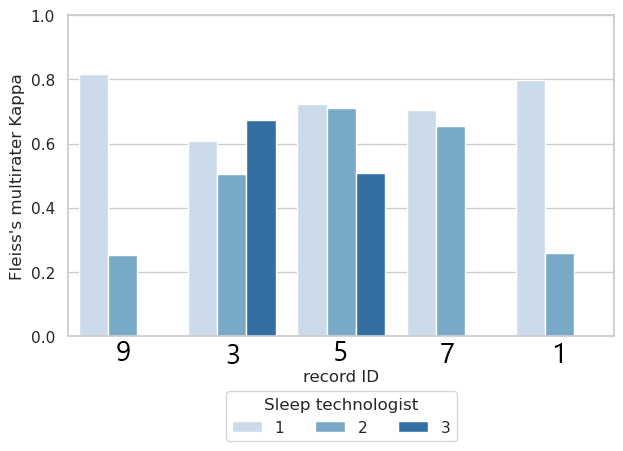}
                        \caption{Fleiss's multiraters $\kappa$ of only grey area epochs with \textbf{ST1} using \asaga assistance.}
                        \label{fig:FKGA1}
                    \end{minipage}
                
                    \setcounter{figure}{4}
                    \setcounter{subfigure}{2}
                    \begin{minipage}[b]{0.445\linewidth}
                        \includegraphics[scale=0.45]{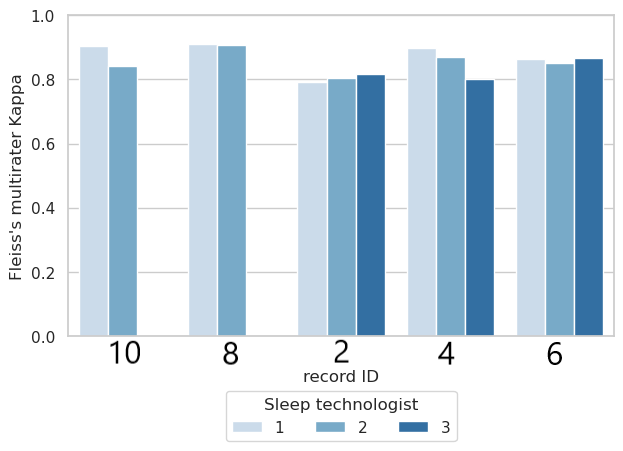}
                        \caption{Fleiss's multiraters $\kappa$ overview with \textbf{ST2} using \asaga assistance.}
                        \label{fig:FKO2}
                    \end{minipage}
                    \hfill
                    \setcounter{figure}{4}
                    \setcounter{subfigure}{3}
                    \begin{minipage}[b]{0.445\linewidth}
                    \includegraphics[scale=0.45]{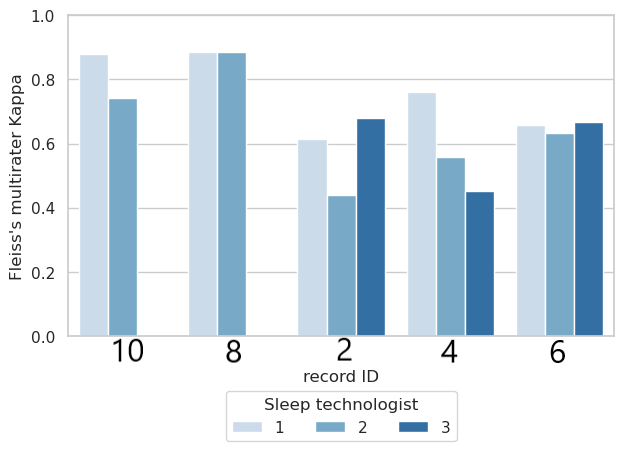}
                        \caption{Fleiss's multiraters $\kappa$ of only grey area epochs with \textbf{ST2} using \asaga assistance.}
                        \label{fig:FKGA2}
                    \end{minipage}

                    \setcounter{figure}{4}
                    \setcounter{subfigure}{4}
                    \begin{minipage}[b]{0.445\linewidth}
                    \includegraphics[scale=0.45]{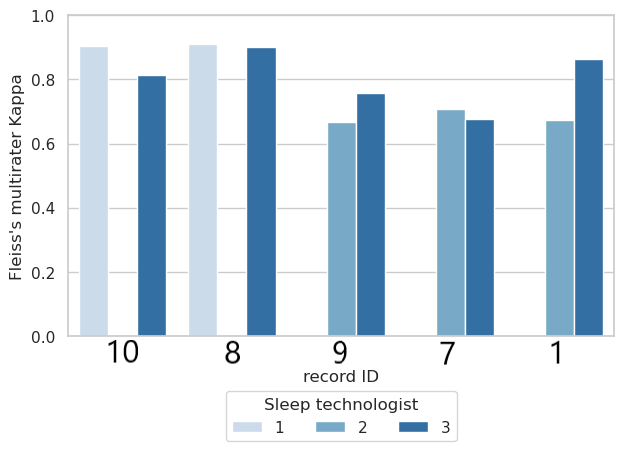}
                        \caption{Fleiss's multiraters $\kappa$ overview with \textbf{ST3} using \asaga assistance.}
                        \label{fig:FKO3}
                    \end{minipage}
                    \hfill
                    \setcounter{figure}{4}
                    \setcounter{subfigure}{5}
                    \begin{minipage}[b]{0.445\linewidth}
                    \includegraphics[scale=0.45]{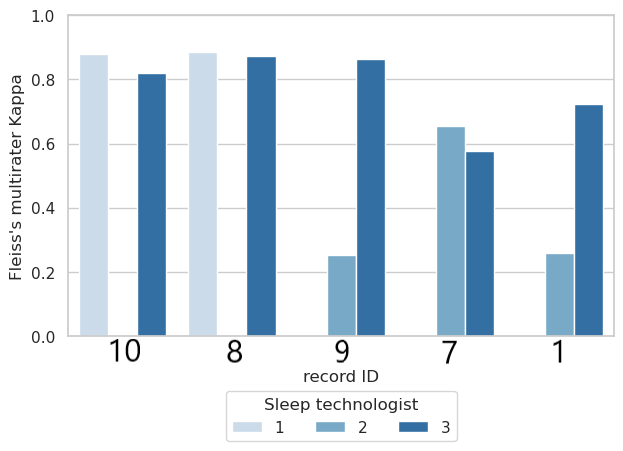}
                        \caption{Fleiss's multiraters $\kappa$ of only grey area epochs with \textbf{ST3} using \asaga assistance.}
                        \label{fig:FKGA3}
                    \end{minipage}
                \setcounter{figure}{4}
                \setcounter{subfigure}{-1}
                    \caption{Agreement analysis of 10 sleep technologists compared to a sleep technologist with or without \asaga assistance. The first column \textbf{(a,c,e)} shows the total agreement per polysomnographs. The second column \textbf{(b,d,f)} shows the agreement for grey area epochs tagged by the artificial intelligence.}
                    \label{fig:kappa}
            \end{subfigure}
            
            Figure \ref{fig:kappa} shows the agreement analysis of each of the three sleep technologists in this study with the $50\times 10$ \psg with and without \asaga.
            The performance levels of ST3, represented in Figures \ref{fig:FKO1} and \ref{fig:FKO2}, consistently matched or surpassed the levels achieved by ST1 and ST2 while using the \asaga tool. However, there is a discernible decrease in scoring agreement for the three \psgs illustrated in Figure \ref{fig:FKO3} when ST3 utilizes \asaga.
            Despite this decrease, the performance levels of ST3 generally remain comparable to and occasionally exceedef those of ST2. 
            Moreover, across all three figures (\ref{fig:FKO1}, \ref{fig:FKO2} and \ref{fig:FKO3}), the agreement of the sleep technologists stayed consistent, indicating that the scorings produced traditionally and using \asaga are comparable.
            For instance, \psgs IDs 1, 7, 8, and 10 all display complete agreement, having been scored twice by the sleep technologists using \asaga.

            A second aspect depicted in Table \ref{TAB:KAPPA} is the agreement of the sleep technologists only calculated for the gray areas on both samples (sleep technologists using \asaga assistance and without \asaga assistance).  
            In this table, ST2 and ST3 got a marked increase in their agreement when using \asaga, with ST2 and ST3 gaining approximately 0.17 $\kappa$, but with ST1 a decrease of 0.03 $\kappa$ for the gray areas.  
            
            Figures \ref{fig:FKGA1}, \ref{fig:FKGA2} and \ref{fig:FKGA3} offer the Fleiss's multiraters $\kappa$ estimated on only gray areas per \psg ID. In this comparison, only the sleep technologist using \asaga knew that these epochs were labeled as gray areas. As expected, in Figures \ref{fig:FKGA1}, \ref{fig:FKGA2}, and \ref{fig:FKGA3}, ST1 depicted the same stability observed previously. ST2 and ST3 showed an increase in $\kappa$ when using \asaga. 
            The increase, however, was less strong for ST2 who showed more agreement's dispersion among the \psgs. Otherwise, in Figure \ref{fig:FKGA3}, ST3's agreement showed a significant increase for \psgs 1 and 9 compared to ST2's agreement which might explain the difference in Figure \ref{fig:FKO3}. This last result indicates that \asaga assistance may benefit a beginner more than an experienced sleep technologist.
            
            In summary, all the participating sleep technologists showed a decrease in their time to score but to a different degree. Regarding their scoring agreement, the sleep technologists depicted three distinct results when using \asaga.
            The agreement of the experienced sleep technologist with the \psg signals was not affected by \asaga. On the other hand, the second experienced sleep technologist has shown more dispersion among the \psg with on average an increase when using \asaga. Finally, the third, less experienced sleep technologist benefited the most from \asaga assistance.

        \subsection*{Interviews with Sleep Technologists}

\begin{figure}[h]
                \centering
                \includegraphics[width=0.6\linewidth]{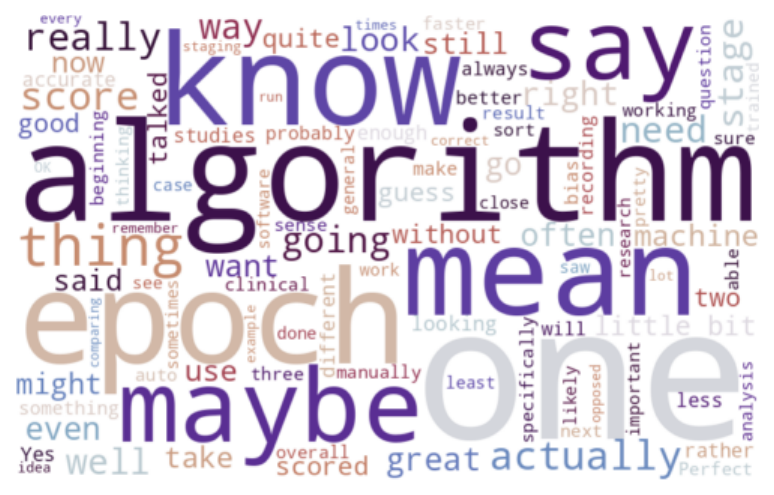}
                \caption{A word cloud of the interview transcripts.}
                \label{fig:wordcloud}
            \end{figure}

            The sleep technologists were interviewed about their experience when using \asaga. The transcript of the interviews was compiled as a word cloud (Figure \ref{fig:wordcloud}).
            
            Initially, the sleep technologists approached the new system with optimism. ST1 expressed initial enthusiasm: "[Before starting] I was very optimistic that it would decrease the scoring time". All sleep technologists found it simple to integrate AI scoring with gray areas into their current workflows with ST3 commenting "I do not think it is an issue at all [...] it is pretty easy to implement".
    
            However, as they used the new system, the sleep technologists noticed a need for a more accurate staging algorithm, with ST1 noting "What I saw is that the algorithm is not good enough". For ST1 and ST3, improved accuracy is essential for reducing the scoring time and building trust in the new system. ST2 provided a slightly different perspective, suggesting that the system's staging accuracy might already be on par with the inter-scorer agreement of human sleep technologists. 

            Overall, all three sleep technologists expressed in various ways that trust in AI technology is significant for its continual adaptation into their practice.    
            The sleep technologists articulated the psychological impact of integrating AI staging with gray areas into their workflows. 
            ST2 expressed concern that the AI suggestion might slightly shift their bias in selecting a sleep stage. ST3 spoke along similar lines: "Maybe I had an unconscious bias to lean towards the [suggested sleep stage]".

            Overall, the sleep technologists found the new system promising and were optimistic about the approach. ST3 was interested in seeing a more detailed quantification of the gray area uncertainty, asking for "the percentage of the prediction or something like that". 
            However, all sleep technologists agreed that improving the staging algorithm's accuracy was important, as ST1 put it: "You need to trust the algorithm".
            Their sentiments reflected a cautious optimism, recognizing the potential benefits while anticipating enhancements in usability and trust as the accuracy of the underlying staging algorithm improves.

\section*{Discussion}

    In the present paper, an advanced web platform has been introduced to fill the gap of sharing, processing, and storing three nights of \psg in the sleep field. The developed platform is built into three distinct components: a front end, a \psg splitter, and a processor component with automatic scoring and storing of each \psg. 
    The front end is connected with the two subsequent parts using a flexible message-queue protocol, preventing the front end from crashing in case of failure in the processing of \psgs. The platform was tested on a set of 60 three-night \psgs files. It showed the average processing time of the platform ranged between 5.6 min an associated file size of 1920 Mo and 7.6 min for 2400 Mo.

    Moreover, the automatic scoring, including the gray areas implemented in the processor component has been assessed with the help of three sleep technologists. 
    The predicted scores by the platform showed a decidedly positive effect on the speed of scoring. This enhancement is achieved without significantly complicating the workflow of sleep technologists. The strategic incorporation of AI support into their routine not only optimizes the time efficiency of scoring but also adds a layer of precision and reliability to the process. The most experienced sleep technologists showed a high agreement on an average of 0.85 $\kappa$ when using AI support. This value of the agreement is in line with the observed agreement obtained for other data sets manually scored \citep{doi:10.5664/jcsm.2350}. 
    Additionally, a significant increase in both the scoring speed and agreement was observed for the less experienced sleep technologist, suggesting that the use of automatic algorithms and gray area assistance has the potential to bridge the gap between more experienced sleep technologists and the less experienced ones, and thus speeding up the training of new sleep technologists.

    \subsection*{Platform Insights}
        Utilizing a message queue protocol imparted a considerable complication in implementing the platform that would have been avoidable if we had instead opted for a separate process using e.g. HTTP requests, or implementing the splitting and processing as part of the same program as the front end. Utilizing message queues in favor of more ad-hoc solutions allowed us more flexibility and scalability than with other solutions.
        The need to split \psgs similarly complicated the work, since it added a component to the process. However, the benefits gained from working with separate nights later in the process outweighed this added complexity.

        Although the web platform was architectured with the express purpose in mind for it to be scalable and robust, this paper does not include an extensive scalability evaluation of the web platform itself. In the current study, this was not the focus, as the platform was tested and evaluated primarily on the improvement it could provide in the task of scoring \psg. Future works could be directed toward stress-testing the platform, evaluating the maximum number of \psg it can handle simultaneously, and determining whether the web platform could sustain heavy traffic loads without considerably slowing down or crashing.
        The period from when a \psg is uploaded to the point it was prepared for the sleep technologist to start scoring has been quantified and documented.
        Since neither the splitter nor processor queue grew to prohibitively big lengths during testing, we did not see a reason to implement scaling functions, nevertheless, the implementation of the system as a whole lends itself well to dynamic scaling.

        In the results part, the processor component has been evaluated over a study composed of three sleep technologists with different experiences scoring 10 \psgs with and without \asaga. 
        Only global metrics such as the scoring duration and the agreement of the sleep technologists have been considered in this paper. 
        These metrics helped to conclude major concerns about AI assistance in manual scoring showing a decrease in scoring time without a significant effect on the agreement between sleep technologists.
        However, this study does not go into detail about the source of the uncertainty in sleep staging between sleep technologists. 
        For instance, it is well known that one primary uncertainty source is the transition between the sleep stages N2 and N3 \citep{jouan2023algorithmic,bakker2023scoring}. 
        Studying if this uncertainty is still encountered after the gray areas revision would be interesting. 
        Moreover, the dispersion obtained in the results reflected a lack of \psg required to obtain an accurate representation of the time to score and sleep technologist's agreement distributions. A study with a greater number of \psgs would allow us to validate the result obtained in the presented paper.

        Using \asaga, the effectiveness of the sleep technologists in terms of scoring duration is affected differently. Their disparity may be explained by the difference in experience with the self-applied PSG frontal signals, the baseline speed of both sleep technologists, and the trust given to the AI-predicted scores in the gray areas. Moreover, a study with more \psgs and sleep technologists is needed to have a better estimation of the effectiveness obtained by the use of AI as a scoring support tool.
        
        As Table \ref{TAB:KAPPA} shows, ST2 and ST3 achieved a significant increase in consensus when scoring with \asaga, while ST1 observed a slight decrease. The decrease for ST1 may be attributed to the low number of epochs that required processing, and the signals being difficult to interpret.
        This tool offers a steady benchmark for scoring, effectively reducing discrepancies between the sleep technologists' results.

        The sleep technologists agreed that the platform integrated well into their workflow, with ST3 commenting especially on the ease of implementation.
        The sleep technologists did raise issues with the performance of the scoring algorithm itself, with ST1 reporting that the scoring algorithm is "not good enough".  ST3 expressed some concern that the sleep stage recommendation system was influencing their decision-making. This worry reflects the need for trust and alignment between the sleep technologist and the algorithms, especially in the context of healthcare AI recommendation systems.
        As the final sentiment of ST3 indicates, the experts display interest in having more insight into the reason why the algorithm assigned areas as gray, aligning with the rise in demand for xAI, reflecting a broader desire for transparency and clarity in human-in-the-loop AI systems. 

        \subsection*{Clinical Acquiescence of AI}
        Traditional accuracy and agreement measures are both derived from the confusion matrix offering an overview of the performance of the classification algorithm. Accuracy variation across different datasets from less than 1\% is pictured as insignificant for that kind of algorithm \citep{rusanen2023asaga,phan2023seqsleepnet}. However, confusion matrix derived metrics such as accuracy only assess the algorithm prediction matching the correct output. It does not guarantee that the algorithm captures a key signal pattern related to a specific sleep stage hiding in this 1\% accuracy variation. For clinical experts, such as sleep technologists, it is crucial to ensure that key signal patterns are correctly interpreted. If a scoring algorithm with high accuracy and agreement is missing these key patterns, it becomes hard for the sleep technologist to trust the algorithm's prediction. 
        To resume, there is a need for a metric assessing the scoring algorithm's conformity that also assures sleep technologists' trust in the algorithm. Clinical acumen is a term symbolizing the ability of healthcare professionals to make quick and accurate decisions on complex issues that a clinical AI along with a human-in-the-loop might include in the future to make a diagnosis \citep{krause2018grader}. In this work, we would like to introduce a general term to define the act of accepting or agreeing to the use of AI as a decision-making tool by clinical experts: Clinical Acquiescence.

    \subsection*{Future Work}
        In the future, a replication of this study needs to be performed, with a greater number of both sleep technologists and a larger subset of \psg to gain a broader perspective of the effects of integrating AI augmentation into the sleep technologist's workflow, along with algorithm trust assessment.

        The next step would be to loop the manual review of the gray area with the automatic scoring algorithm. 
        This process is referred to as active learning \citep{settles2009active,ren2021survey}, and aligns with the AI-integrated human-in-the-loop workflow. 
        A continuous loop would link the reviewed gray area with the scoring AI updating the model and sending a new set of gray areas corresponding to the actual sleep technologist.

        Due to the modularity of the platform, it is easy to add more algorithms and augments to the processor, making the adoption of any additional algorithms more approachable without resulting in downtime or causing data loss. For example, the  BreathFinder \citep{Holm_BreathFinder_2020} respiratory isolation algorithm is planned for addition to the processor to allow future analysis of individual respiratory cycles. Additionally, adding new destinations and output formats for the \psgs is made easy, e.g. usinga micro-scoring platform with integrated machine-learning capabilities, currently under development.
        
        One possible avenue to further advance the platform is to allow researchers to upload their custom automatic scoring algorithms to be vetted and be run autonomously on test data, without ever having to gain physical or digital access to the data, allowing for a reliable method for testing disparate algorithms on the same datasets for greater consistency, reproducibility and transparency in future sleep research.

\section*{Conclusion}

    We observed a clear gap in research addressing the integration and evaluation of automatic scoring algorithms for \psg.
    In this work, we presented a platform that enables \psg collection, integrated with automatic AI scoring algorithms. We evaluated the platform in terms of its effect on sleep technologists' time, and accuracy when scoring \psgs that incorporate AI assistance. 

    The proposed platform incorporates AI assistance but still prioritizes the human expert as the ultimate decision-maker.
    This balance of human expertise and AI presents a promising avenue for future advancements in the field of sleep study and analysis, potentially leading to more refined and accurate diagnostic practices.

\section*{Conflict of Interest Statement}
    E.S. Arnardottir  reports lecture honoraria from Nox Medical, Philips, ResMed, Jazz Pharmaceuticals, Linde Healthcare, Wink Sleep, Vistor (NovoNordisk) and Apnimed, participation on the Philips Sleep Medicine and Innovation medical advisory board, and has a leadership role as Secretary of the board of the European Sleep Research Society, outside the submitted work.

\section*{Author Contributions}

BH: Conceptualization, Data curation, formal analysis, Investigation, Methodology, Project administration, Resources, Visualization, Writing – original draft;
GJ: Conceptualization, Data curation, formal analysis, Investigation, Methodology, Project administration, Software, Visualization, Writing – original draft;
EH: Conceptualization, Data curation, formal analysis, Investigation, Methodology, Software, Visualization, Writing – original draft;
SS: Conceptualization, Data curation, Investigation, Methodology, Resources, Writing – original draft;
KH:, Data curation,  Writing – original draft;
CM:, Data curation,  Writing – review \& editing;
ESA: Conceptualization, Funding acquisition, Methodology, Resources, Supervision,  Writing – review \& editing;
MÓ: Conceptualization, Funding acquisition, Investigation, Methodology, Project administration, Resources, Supervision,  Writing – original draft;
ASI: Conceptualization, Funding acquisition, Investigation, Methodology, Project administration, Resources, Supervision,  Writing – original draft.
    
\section*{Acknowledgments}
    We express our gratitude to Nox Medical for their essential software support in our research. Their tools and expertise significantly contributed to our work. We appreciate their commitment and support, without which the work described in this paper would have been impossible.

    The authors of this paper have received funding to craft this paper from the European Union’s Horizon 2020 research and innovation programme (grant agreement 965417) as well as NordForsk (NordSleep project 90458) via Business Finland (5133/31/2018), the Icelandic Research Fund (ESA \& ASI).

    The Sleep Revolution project has received funding from the European Union’s Horizon 2020 research and innovation program under grant agreement No 965417. 
    The sleep technologists who scored the studies--- Heidur Grétarsdóttir, Marjo Sunnari, Beate Diecker, Jacob Siegert, Dina Fernandes, Cátia Lígia Rito de Oliveira, Elena Robbi, Paul Murphy, and Alexander Ryan---are especially thanked for there essential contribution to this paper, as well as Kristín Anna {\'O}lasfdóttir, who led the team of scorers.
    
\section*{Data Availability Statement}
    Due to the datasets used, and generated by this work being confidential medical data, they cannot be made public in any form.

\section*{Code Availability Statement}
    All code has been made publicly available and open source via GitHub code repositories. 
    The frontend platform can be accessed via \href{https://github.com/Sleep-Revolution/sleepscorerapi}{https://github.com/Sleep-Revolution/sleepscorerapi}, the splitter service can be accessed via \href{https://github.com/Sleep-Revolution/ESADASplitterService}{https://github.com/Sleep-Revolution/ESADASplitterService}, and finally, the processor can be accessed via \href{https://github.com/Sleep-Revolution/EsadaFileConsumer}{https://github.com/Sleep-Revolution/EsadaFileConsumer}.

\bibliographystyle{Frontiers-Harvard} 
\bibliography{references}

\end{document}